\documentclass[%
reprint,
 superscriptaddress,
 amsmath,amssymb,
prb,
floatfix,
]{revtex4-2}

\usepackage{graphicx}
\usepackage{dcolumn}
\usepackage{bm}
\usepackage{amsmath}
\usepackage{comment}
\usepackage{appendix}
\usepackage{ulem}

\usepackage[colorlinks,linkcolor=blue,citecolor=blue,anchorcolor=blue]{hyperref}

\begin{document}

\title{Visualizing Vortex Cluster Dynamics in the Weak Type-II Superconductor CaSb$_2$}

\author{Yusuke Iguchi}
\email{yiguchi.phys@gmail.com}
\affiliation{Stanford Institute for Materials and Energy Sciences, SLAC National Accelerator Laboratory, 2575 Sand Hill Road, Menlo Park, California 94025, USA}
\affiliation{Geballe Laboratory for Advanced Materials, Stanford University, Stanford, California 94305, USA}

\author{Nabhanila Nandi}
\affiliation{Stanford Institute for Materials and Energy Sciences, SLAC National Accelerator Laboratory, 2575 Sand Hill Road, Menlo Park, California 94025, USA}
\affiliation{Geballe Laboratory for Advanced Materials, Stanford University, Stanford, California 94305, USA}

\author{Mohamed Oudah}
\affiliation{Stewart Blusson Quantum Matter Institute, University of British Columbia, Vancouver, British Columbia V6T 1Z4, Canada}
\affiliation{Department of Physics and Astronomy, University of British Columbia, Vancouver, Canada V6T 1Z1, Canada}

\date{\today}

\begin{abstract}
Scanning SQUID imaging of CaSb$_2$ reveals dense vortex clusters with enhanced boundary susceptibility and suppressed internal vortex motion, which features inconsistent with both isolated‑vortex and flux tube behaviors. These measurements provide the first local visualization of magnetic dynamics within vortex clusters in a weakly pinned superconductor, offering a new route to probe non‑monotonic vortex–vortex interactions that are typically expected in single‑band type‑II/1 or multiband type‑1.5 superconductors. Although the superfluid density follows a single‑gap BCS model and the Ginzburg–Landau parameter of CaSb$_2$ lies slightly outside the type‑II/1 regime, vortex clustering and spatially inhomogeneous dynamics are clearly observed, indicating physics beyond existing microscopic theories for single‑band superconductors.
\end{abstract}


\maketitle



The spatial arrangement of vortices in superconductors reflects the interplay of magnetic screening, superfluid stiffness, and vortex–vortex interactions. While type-II superconductors typically host a repulsively interacting Abrikosov vortex lattice~\cite{Abrikosov1957}, theoretical studies predict non-monotonic interactions, with short-range repulsion and long-range attraction, near the critical Ginzburg–Landau parameter $\kappa_0 = 1/\sqrt{2}$~\cite{Eilenberger1969, Brandt1976, Klein1987}. Microscopic calculations based on the Eilenberger or Bogoliubov–de Gennes formalisms confirm that clean single-component superconductors with intermediate $\kappa$ values ($0.6 \lesssim \kappa \lesssim 1.1$ at $T=0$) can support attractive long-range interactions and form vortex clusters, hallmarks of the type-II/1 regime\cite{Klein1987, Weber1987, Neverov2024}. 
Beyond the single-component case, non-monotonic vortex–vortex interactions can also emerge in multicomponent or symmetry-broken superconductors through distinct microscopic mechanisms, including the type-1.5 regime in multiband systems~\cite{Babaev2005, Silaev2011, Carlstrom2011} and related effects in systems with multilayer structures~\cite{Varney2013}, strong anisotropy~\cite{Silaev2018}, coexisting magnetic orders~\cite{Tachiki1979}, or non-centrosymmetric pairing~\cite{Garaud2020, Samoilenka2020}.
Experimentally, vortex clusters and related morphologies have been directly imaged in several candidate materials~\cite{Huebener2001,Brandt-Essmann1987,Moshchalkov2009, Nishio2010,Ge2014,Ooi2025}.
However, their origin remain debated: clustered configurations may reflect thermodynamically stable states arising from intrinsic non-monotonic interactions, or result from disorder-induced pinning. Thus, although static magnetic imaging has revealed clustered morphologies, direct experimental access to the underlying vortex interactions remain limited.

CaSb$_2$ is a recently discovered superconductor that provides a promising platform to address this challenge. It crystallizes in a non-symmorphic structure ($P2_1/m$) and hosts topologically non-trivial Dirac-like bands near the Fermi level~\cite{Oudah2022, Ikeda2022}. Evidence for multiband superconductivity comes from penetration depth and specific heat studies~\cite{Duan2022}, while nuclear quadrupole resonance suggests a fully gapped behavior~\cite{Takahashi2021}, leaving the gap structure unresolved. Muon spin relaxation experiments further report time-reversal symmetry breaking~\cite{Oudah2024}. Estimates of $\kappa$ along the $c$-axis range from $\sim 0.95$ to $1.4$, overlapping with the type-II/1 regime, although values within the $ab$ plane are larger ($\kappa \sim 3$)~\cite{Ikeda2020, Ikeda2022, Oudah2022, Oudah2024}. While CaSb$_2$ preserves global inversion symmetry, its non-symmorphic structure implies local inversion symmetry breaking~\cite{Fischer2023}. However its possible consequences for vortex physics remain unclear.
Despite these suggestive parameters, no direct magnetic imaging of vortex configurations in CaSb$_2$ has been reported. In particular, whether the vortex–vortex interactions exhibit non-monotonic behavior is an open experimental question that we address in this work.

\begin{table}[bt]
\centering
\caption{ GL parameters of CaSb$_2$. The GL coherence length $\xi$ and penetration depth $\lambda$ are listed with the crystal directions parallel to the applied field. ($\kappa_0\simeq0.707$)}
\begin{tabular}{c l l l c}
\hline
& $\xi(0)$ [nm] & $\lambda(0)$ [nm] & $\kappa=\lambda/\xi$ & Ref \\ \hline
poly & 36–44 & $>240$ & $>5.4$ & \cite{Ikeda2020} \\ 
$ab$ & 116 & 89.5 & 0.77(.04) & \cite{Oudah2022} \\
$ab$ & 149-166 & & 3.2-3.9 & \cite{Ikeda2022} \\
$ab$ & 202 & & 2.90(.26) & \cite{Oudah2024} \\
$c$ & 52-57 & & 1.1-1.4 & \cite{Ikeda2022} \\
$c$ & 66 & & 0.95(.12) & \cite{Oudah2024} \\ 
$c$ &  & 93 & 1.41–1.79 & \cite{Duan2022} \\ 
$c$ &  & 87-170 & 1.32–3.27 & This work \\ \hline
\end{tabular}
\end{table}

\begin{figure}[tb]
\begin{center}
\includegraphics*[width=7cm]{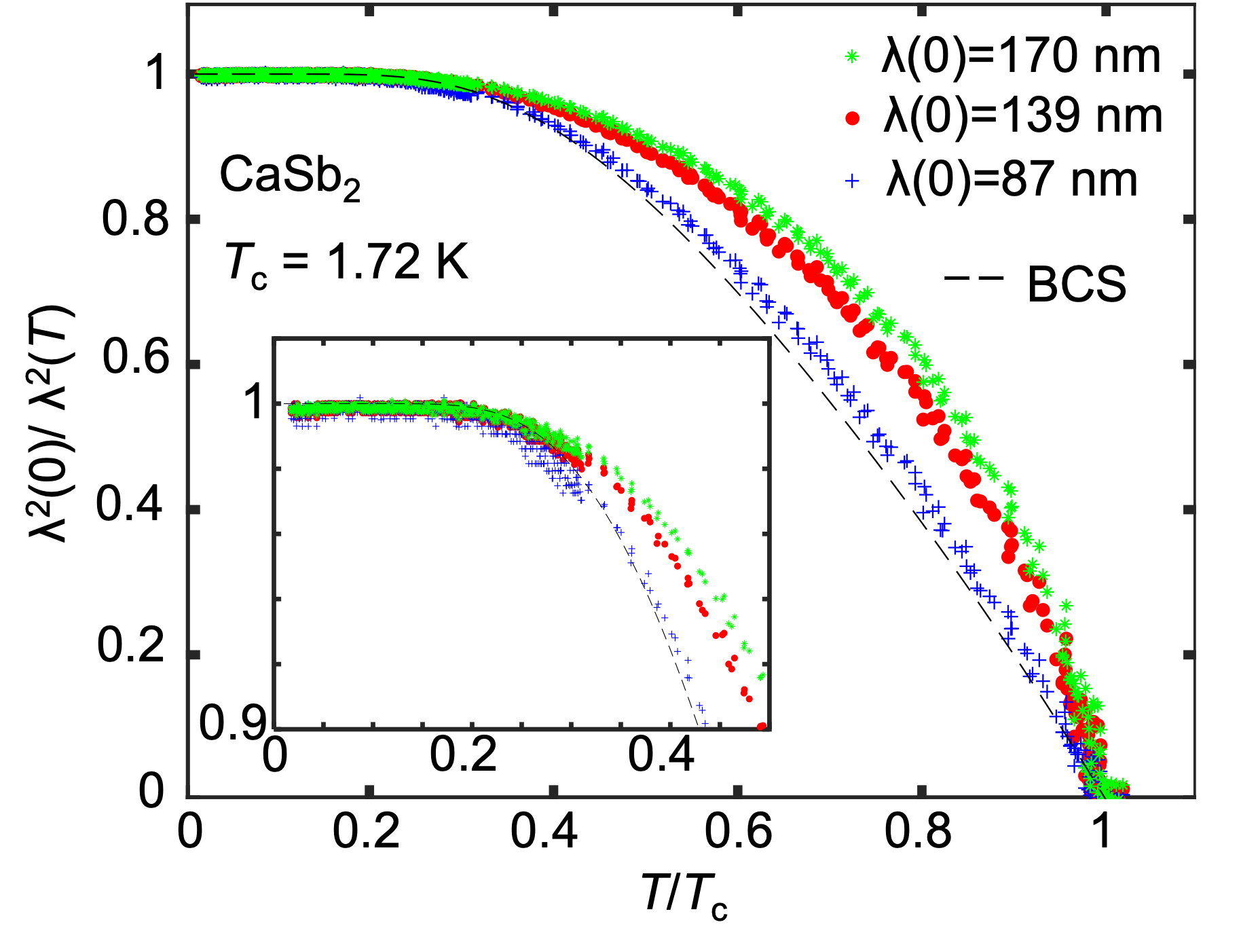}
\caption{\label{fig:A1} 
Scanning SQUID susceptometry measurements on CaSb$_2$ with the applied field normal to $ab$ plane yield a temperature dependence of the superfluid density consistent with a single-gap BCS model, showing clear deviations from previously proposed multiband models~\cite{Duan2022}. 
}
\end{center}
\end{figure}

\begin{figure*}[t]
\begin{center}
\includegraphics*[width=16cm]{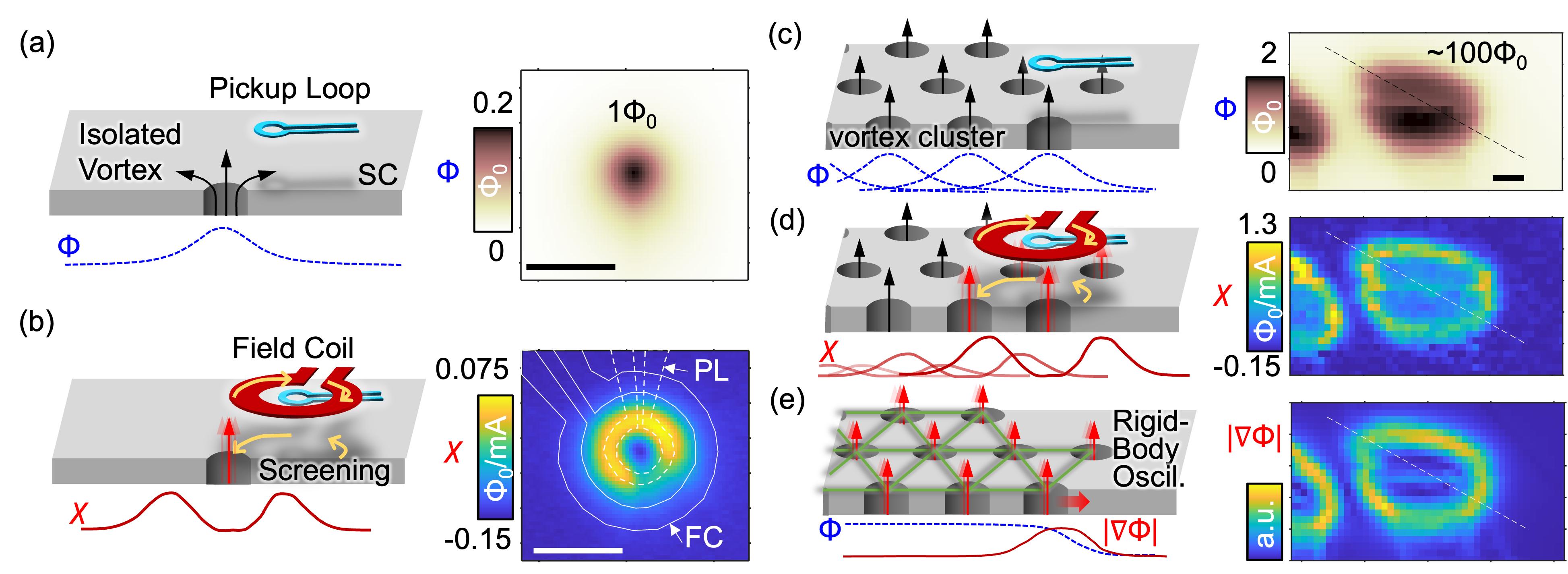}
\caption{\label{fig:cluster} Local observation of partially collective vortex cluster dynamics.
(a-b) Magnetometry $\Phi$ and susceptometry $\chi$ over an isolated vortex at 1.5 K in CaSb$_2$, together with schematic illustrations (left). These serve as a reference for evaluating cluster behavior.
(c–e) Left: schematic images of (c) magnetometry over a vortex cluster, (d) locally induced magnetic dynamics in a cluster, and (e) rigid-body-like collective oscillation of the entire cluster (shown for comparison). Right: corresponding measurements, (c) flux scan (from Fig.~3(c)), (d) susceptibility scan (from Fig.~3(g)), and (e) the derivative $|\nabla\Phi|$ calculated from (c), representing the rigid-body approximation mode. Scale bars: 5~\textmu m. 
}
\end{center}
\end{figure*}

In this work, we present a new application of scanning superconducting quantum interference device (SQUID) susceptometry, probing the dynamic magnetic response of vortices within clusters. Using this method on the weak type-II superconductor CaSb$_2$, we simultaneously image static flux and dynamic local susceptibility, uncovering vortex clustering with enhanced boundary response and suppressed internal dynamics. These features are suggestive of non-monotonic vortex–vortex interactions. Although low pinning in CaSb$_2$ prevents direct extraction of the intrinsic superfluid density within a cluster, our results establish the viability of this approach and define the experimental conditions necessary to differentiate between type-1.5 and type-II/1 superconductivity.

Single crystals of CaSb$_2$ were grown using an Sb self-flux method~\cite{Oudah2022}. We used a scanning SQUID susceptometer consisting of a pickup loop (PL) and a concentric field coil (FC) (see outlines in Fig.~2(b))~\cite{Kirtleyrsi2016}. The PL measures the local magnetic flux $\Phi$ in units of the flux quantum $\Phi_0$ at a loop–sample separation $z$, referred to as the height. An ac current of $|I^{\text{ac}}| =$ 0.004–0.1~mA at 1020~Hz, supplied by an SR830 Lock-in-Amplifier, was applied through the FC to generate a localized oscillating magnetic field. Maximum fields generated at the sample surface is estimated as 0.5–13.4~\textmu T at the typical scan height 1150~nm. We recorded both the quasi-static flux $\Phi$ and the ac flux response
$\Phi^{\text{ac}}$ over the $ab$-plane surface of as-grown CaSb$_2$, and express the local ac susceptibility as $\chi = \Phi^{\text{ac}} / |I^{\text{ac}}|$ in units of $\Phi_0$/A.

To establish the fundamental superconducting properties of our CaSb$_2$ samples, we first performed scanning SQUID susceptibility measurements as a function of temperature ($T = 0.03$–1.74~K) on the same crystal used for vortex imaging. As shown in Fig.~1, the superconducting transition was clearly identified at $T_\text{c} = 1.72$~K. From the local susceptibility signal, we estimated the magnetic penetration depth $\lambda(T)$ with an effective radius of 790~nm and scan height of 400~nm ($\pm$30~nm) by following the established method~\cite{Kirtley2012}. The extracted zero-temperature penetration depth $\lambda(0)$ ranged from 87 to 170~nm, consistent with earlier tunneling diode oscillator measurements~\cite{Duan2022}. However, the temperature dependence of the normalized superfluid density $\lambda^2(0)/\lambda^2(T)$ closely follows a single-band fully-gapped Bardeen-Cooper-Schrieffer (BCS) model with $\lambda(0) = 87$~nm, in contrast to previous claims of multiband superconductivity~\cite{Duan2022}. 

We then established the characteristic response of an isolated vortex in CaSb$_2$ by imaging in magnetometry and susceptometry modes [Figs.~2(a) \& 2(b)]. These measurements serve as the reference for evaluating the behavior of vortex clusters. The magnetic flux profile of the isolated vortex at 1.5~K [Fig.~2(a)] fits well the monopole model~\cite{Kirtleysst2016} [Fig.~S1]. The best-fit penetration depth is 210~nm, consistent with the susceptibility measurements that estimated 170–250~nm at 1.5~K. The paramagnetic signal surrounded by the vortex center in Fig.~2(b) is the typical vortex dynamics induced by the applied field from the FC in isotropic pinning potentials~\cite{Irene2019,Iguchi2021}. 

After 1.7~Oe uniform field cooling, we observe a localized flux structure with relatively uniform interior [Figs.~2(c)]. The enclosed flux within the cluster is $\sim105\Phi_0$, calculated by convoluting the magnetic field with SQUID's point spread function (for details, see Fig.~S3 and Ref.~\cite{SpantonThesis}). While our spatial resolution does not resolve individual vortices, the local susceptibility profile [Fig.~2(d)] exhibits sharply enhanced peaks along the boundary and suppressed signal inside. This spatial pattern, distinct from either isolated vortices or homogeneous flux penetration, serves as the starting point for our dynamical analysis.

\begin{figure}[tb]
\begin{center}
\includegraphics*[width=8.5cm]{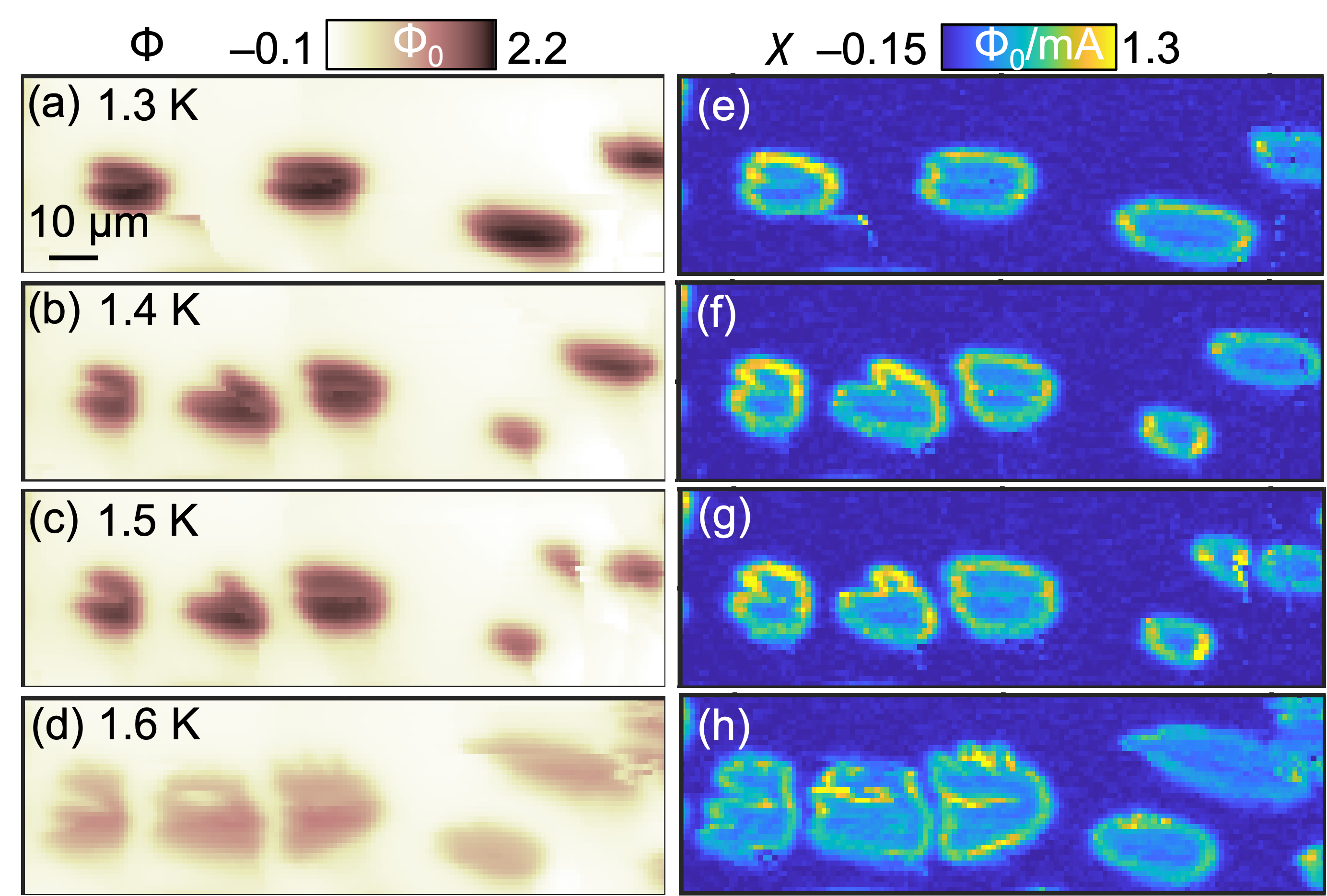}
\caption{\label{fig:T-dep} Temperature-induced morphological transformations and consistent dynamic magnetic responses of vortex clusters in CaSb$_2$.
(a-d) Magnetometry and (e-h) susceptometry of vortex clusters formed under under 1.7~Oe field cooling, shown at temperatures increasing from 1.3~K to 1.6~K. While the cluster morphology evolves from compact to elongated or fragmented shapes with increasing temperature, the enhanced boundary response persists, indicating robust collective vortex dynamics.
}
\end{center}
\end{figure}

Temperature-dependent magnetometry measurements [Figs.~3(a-d) and the derivatives in Fig.~S5] show that the cluster evolves from compact to finger-like shapes as $T$ increases toward $T_\text{c}$, while the susceptibility profile [Figs.~3(e-h)] maintains sharp boundary peaks and similar intra-cluster response. Together with the several field-cooled measurements shown in Fig.~S6, this suggests that vortex clusters in CaSb$_2$ retain similar dynamic structure over a finite thermal range.

\begin{figure}[tb]
\begin{center}
\includegraphics*[width=8cm]{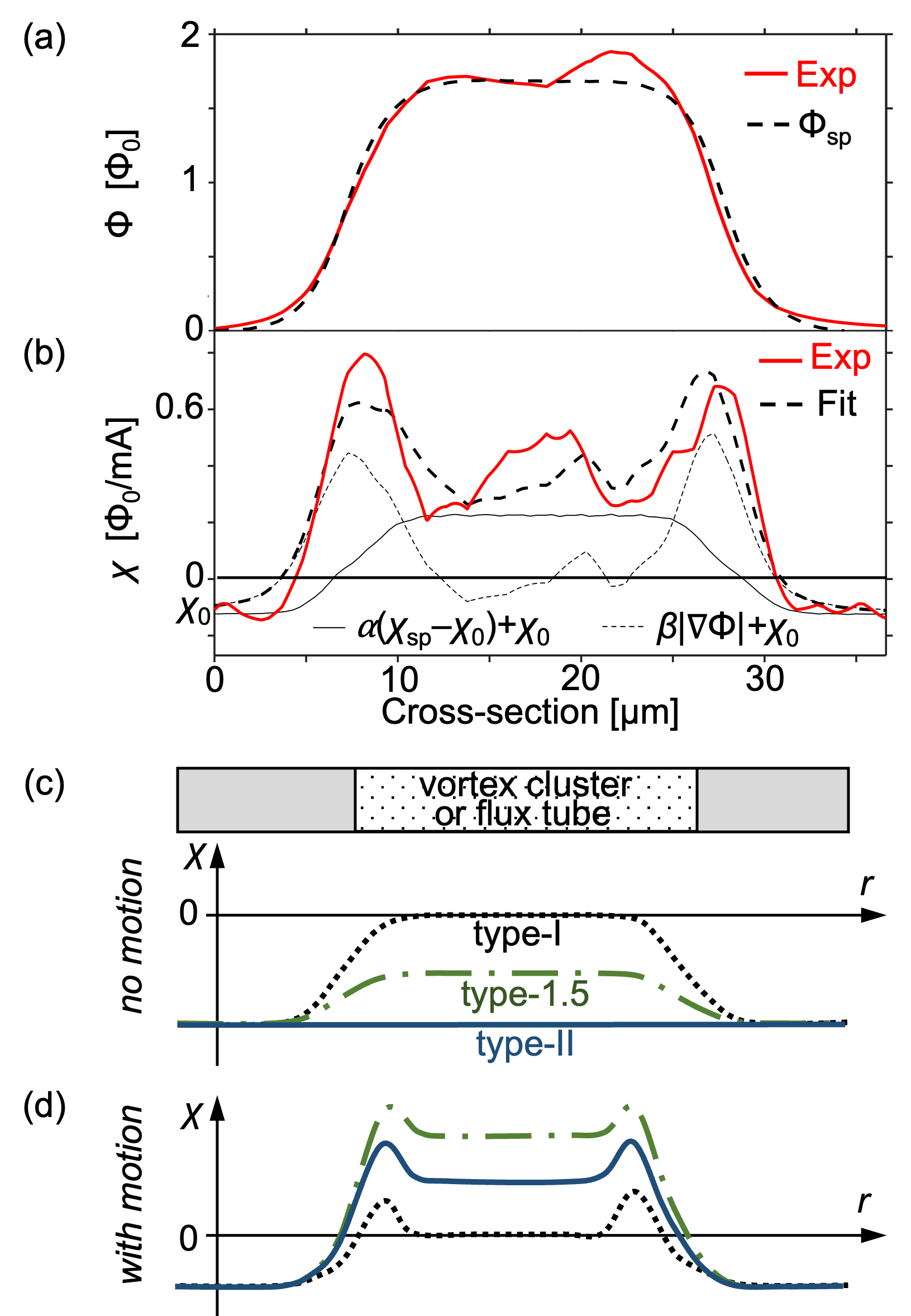}
\caption{\label{fig:Cross} 
Quantitative modeling of vortex clusters and dynamics.
(a) Cross-section of $\phi$ along the dashed line in Fig.~2(c) compared with a triangular-lattice cluster simulation $\Phi_\mathrm{sp}$, constructed by superposing isolated-vortex data with lattice constant $a$ = 1.65~\textmu m (range: 1.55---1.65~ \textmu m).
(b) Cross-section of the measured $\chi$ compared with the scaled model $\alpha(\chi_\mathrm{sp}-\chi_0) + \beta|\nabla \Phi| + \chi_0$, with fit parameters $\chi_0 = -0.12~\Phi_0/\mathrm{mA}$, $\alpha \simeq 0.17$, and $\beta \simeq$ 1.61~\textmu m/mA.  
(c,d) Schematic susceptibility across vortex clusters in various superconductors, contrasting (c) strong-pinning and (d) weak-pinning regimes, with finite internal response appearing only in dynamic type-II and type-1.5 cases.
}
\end{center}
\end{figure}

The static flux profile of the cluster can be reproduced by a simple superposition of isolated-vortex signals, $\Phi_\mathrm{sp}$, arranged in a triangular lattice with lattice constant $a = 1.65$~\textmu m [Fig.~4(a); see also Fig.~S2(a) for the full 2D simulation], while leaving room for alternative lattice configurations with slightly different effective spacing. The realization of such clusters under a weak applied field likely requires an underlying mechanism, such as non-monotonic vortex-vortex interactions or highly inhomogeneous pinning forces, that stabilizes this dense configuration.

In type-I superconducting thin films, flux tubes in the intermediate state exhibit collective dynamics resembling rigid-body motion~\cite{Gech5}. Motivated by this, we simulate the expected cluster response using a rigid-body approximation, assuming the entire structure moves coherently. Such rigid-body-like modes are expected to produce similar susceptibility profiles in both strongly interacting vortex clusters (type-II) and individual flux tubes (type-I). We simulate this response by computing $|\nabla \Phi| = \sqrt{(\partial\Phi/\partial x)^2 + (\partial\Phi/\partial y)^2}$ from the static flux image, assuming a fixed internal structure. The result reproduces the sharp boundary peaks but shows qualitative discrepancies in the interior [Fig.~2(e)].

We next evaluate whether the observed susceptibility can be explained by two limiting cases: a superposition of isolated vortex responses ($\chi_{\mathrm{sp}}$), and a rigid-body cluster response modeled by $|\nabla \Phi|$ [Fig.~4(b)]. While $\chi_{\mathrm{sp}}$ overestimates the interior response, the rigid-body model captures the sharp boundary peaks but fails inside. The best fit suggests an internal suppression factor of $\alpha\sim0.2$, implying strong vortex-vortex binding. These features indicate spatially inhomogeneous dynamics: constrained interior and flexible boundary motion, consistent with collective behavior.

Although the boundary-enhanced signal bears resemblance to rigid-body modes, we confirm that the observed susceptibility cannot be attributed to mechanical vibrations between the sample and SQUID, neither extrinsic nor magnetic-force-induced~\cite{Logan2019}\footnote{See Supplementary Material and Figs.~S4 for detailed estimates of vibration amplitude and comparison with $\partial\Phi/\partial z$.}.

Taken together, the comparison with $\chi_{\mathrm{sp}}$, the rigid-body model, and the vibration estimates shows that the observed susceptibility cannot be explained by independently responding vortices, by coherent rigid translation of the entire cluster, or by sample–sensor vibrations alone. Instead, the data reveal spatially inhomogeneous vortex dynamics: suppressed motion inside the cluster and enhanced response at the boundary. These features point to strong vortex–vortex binding in the interior and weaker, asymmetric interactions near the edge—signatures of collective behavior beyond simple limiting models. Such susceptibility enhancement reflects finite vortex motion under weak pinning, which can occur in both type-II and type-1.5 superconductors [Figs.~4(e,d)]. However, type-I behavior is ruled out by the positive internal susceptibility, as normal domains would yield vanishing response. Strong-pinning systems could show reduced or negative internal signals, but with distinct temperature dependence.

While single-component GL theory supports clustering only in a narrow $\kappa$ window up to $\kappa\sim1.1$, recent BdG studies extend the clustering regime to $\kappa\sim1.4$~\cite{Neverov2024}. Our estimates based on our measured $\lambda$ and estimates of $\xi$ for our crystals~\cite{Oudah2024} give a $\kappa=$1.3–3.3. This places most of the $\kappa$ for CaSb$_2$ beyond the single-component GL and BdG clustering regimes, although the lower bound marginally overlaps with the recent BdG results~\cite{Neverov2024}. Therefore, the observed vortex clustering at $T/T_\text{c}\sim0.8$ still requires further explanation. Combined with a more accurate evaluation of $\kappa$ in future studies will help us determine the origin of vortex clustering in CaSb$_2$.

The nearly uniform static flux combined with spatially inhomogeneous dynamics is qualitatively consistent with multiband superconductivity. However, the temperature dependence of the penetration depth follows a single-gap BCS model, suggesting that if CaSb$_2$ is multiband, the gap amplitudes must be nearly identical. Such conditions may permit type-1.5 behavior if the GL parameters differ across bands, but this requires fine tuning. Our results thus call for new microscopic models that reconcile the presence of non-monotonic interactions with single-gap-like thermodynamics with higher $\kappa$.

Although vortex dynamics obscure direct access to intrinsic superfluid density, they also offer a diagnostic tool: in the strong pinning limit, local susceptibility would reflect the superfluid stiffness [Fig.~4(e)]. In such a regime, type-1.5 superconductors would exhibit reduced diamagnetism inside the cluster~\cite{Carlstrom2011}, while type-II systems would remain uniform. However, in our measurements, finite intra-cluster response persists even at the lowest temperatures, indicating active vortex motion and relatively weak pinning.

This weak pinning is supported by additional observations: vortices were moved during scanning well below $T_\text{c}$, and attempts to reduce scan height using a smaller pickup loop failed due to instability, motivating our use of a larger pickup loop and higher scan height for stability. The high crystalline quality, as evidenced by clear de Haas–van Alphen oscillations and a well-defined Fermi surface~\cite{Oudah2022}, and long coherence length likely suppress atomic-scale pinning, enabling collective vortex dynamics. 

While this weak pinning regime prevents the direct identification of superconducting type via susceptibility-based stiffness measurements, the observed finite intra-cluster response remains a valuable probe of vortex–vortex interaction strength. In particular, the coexistence of suppressed interior dynamics and sharp boundary response encodes information about non-monotonic interactions, which can be extracted through comparison with microscopic models. Thus, even in weakly pinned systems like CaSb$_2$, local susceptibility imaging provides critical insight into the underlying vortex physics and interaction mechanisms.

In summary, we used scanning SQUID imaging to investigate vortex clusters in the weak type-II superconductor CaSb$_2$. The clusters exhibit uniform static flux, enhanced susceptibility at the boundary, and suppressed dynamics inside—indicative of collective behavior driven by non-monotonic interactions. Importantly, this study presents the first local observation of magnetic dynamics within vortex clusters, revealing spatially inhomogeneous responses that encode information about vortex–vortex interaction strength.

While these features resemble type-1.5 superconductivity, the superfluid density follows a single-gap BCS model, suggesting nearly identical gaps across bands. The persistence of intra-cluster dynamics even at low temperatures points to weak pinning, supported by scanning instability and long coherence length. These findings introduce CaSb$_2$ as a new platform for vortex clustering and highlight scanning SQUID susceptometry as a powerful probe of vortex interactions within a cluster. Our results thus call for new microscopic models that can reconcile non-monotonic vortex interactions with single-gap-like thermodynamics in a regime beyond conventional type-II/1 superconductivity.



\begin{acknowledgments}
We thank Kathryn A. Moler and Egor Babaev for fruitful discussion.
This work was primarily supported by the DOE “Quantum Sensing and Quantum Materials” Energy Frontier Research Center under Grant No. DE-SC0021238.
\end{acknowledgments}


\newpage
\clearpage

\setcounter{figure}{0}
\setcounter{equation}{0}
\renewcommand{\thefigure}{S\arabic{figure}}
\renewcommand{\theequation}{S\arabic{equation}}

\onecolumngrid
\appendix

\begin{center}
	\Large
	{Supplemental Material for \\\lq\lq Visualizing Vortex Cluster Dynamics in the Weak Type-II Superconductor CaSb$_2$ \rq\rq} \\by Iguchi $et$ $al.$
\end{center}

\section{scanning SQUID susceptometry and mechanical vibration in a cryostat}
To examine wether the observed susceptibility could arise from mechanical vibrations between the sample and the SQUID, we estimated the effective displacement $\Delta r$ from the fit parameter $\beta$ in Fig.~2(b) via $\chi = |\nabla \Phi| \cdot (\Delta r / |I^{ac}|)$ yielding $\Delta r \sim 64–161$~nm for the range $I^{ac}=0.004–0.1$~mA. These values exceed known cryostat vibrations ($\lesssim20$~nm)~\cite{Logan2019} and critically, exhibit no scaling with $I^{ac}$, ruling out extrinsic vibration sources such as pulse-tube-induced motion.

\section{Magnetic force estimation and vertical displacement due to interaction with vortex cluster fields}
To estimate whether mechanical vibration caused by magnetic interaction between the field coil and vortex cluster fields can explain the observed susceptibility signal, we estimated the vertical magnetic force and resulting displacement of the cantilever, where the SQUID chip is mounted.

We modeled the vortex cluster with the lattice constant of 1.6~\textmu m as a distribution of monopole sources and calculated the magnetic force $F_z$ between the applied ac field gradient $(\partial B_\mathrm{FC}/\partial z)$ and the cluster field $B_\mathrm{mono}$. The monopole fields $B_\mathrm{mono}$ and the applied field $B_\mathrm{FC}$ are calculated by using the scan height of 1.15 \textmu m, the penetration depth of 0.21 \textmu m, and the radius of the field coil of 3.5 \textmu m. For representative applied currents: For $I^{ac}=0.004$~mA, the $z$-component of the magnetic force is $F_z=1.269\times10^{-9}$~N, the vertical displacement is $\Delta z=-2.99\times 10^{-12}$~m. For $I^{ac}=0.1$~mA, the $z$-component of the magnetic force is $F_z=3.174\times10^{-8}$~N, the vertical displacement is $\Delta z=-7.47\times 10^{-11}$~m.

The vertical displacement was calculated from the cantilever's spring constant $k$, estimated using its resonant frequency $f_0=1240$~Hz and the mass of the SQUID chip on the cantilever $m\sim7.0\times10^{-6}$~kg: $k=(2\pi f_0)^2m\sim425$~N/m, $\Delta z = F_z/k$.

Even at the highest applied current, the estimated displacement is much smaller than the measured susceptibility signal would require (typically $\sim$10~nm), and the frequency used (1020~Hz) is far from the cantilever resonance, further suppressing motion. This strongly supports the conclusion that the observed susceptibility is not caused by mechanical vibration due to magnetic forces.





\clearpage
\section*{Supplemental Figures}
    
\begin{figure}[htb]
\includegraphics*[width=12cm]{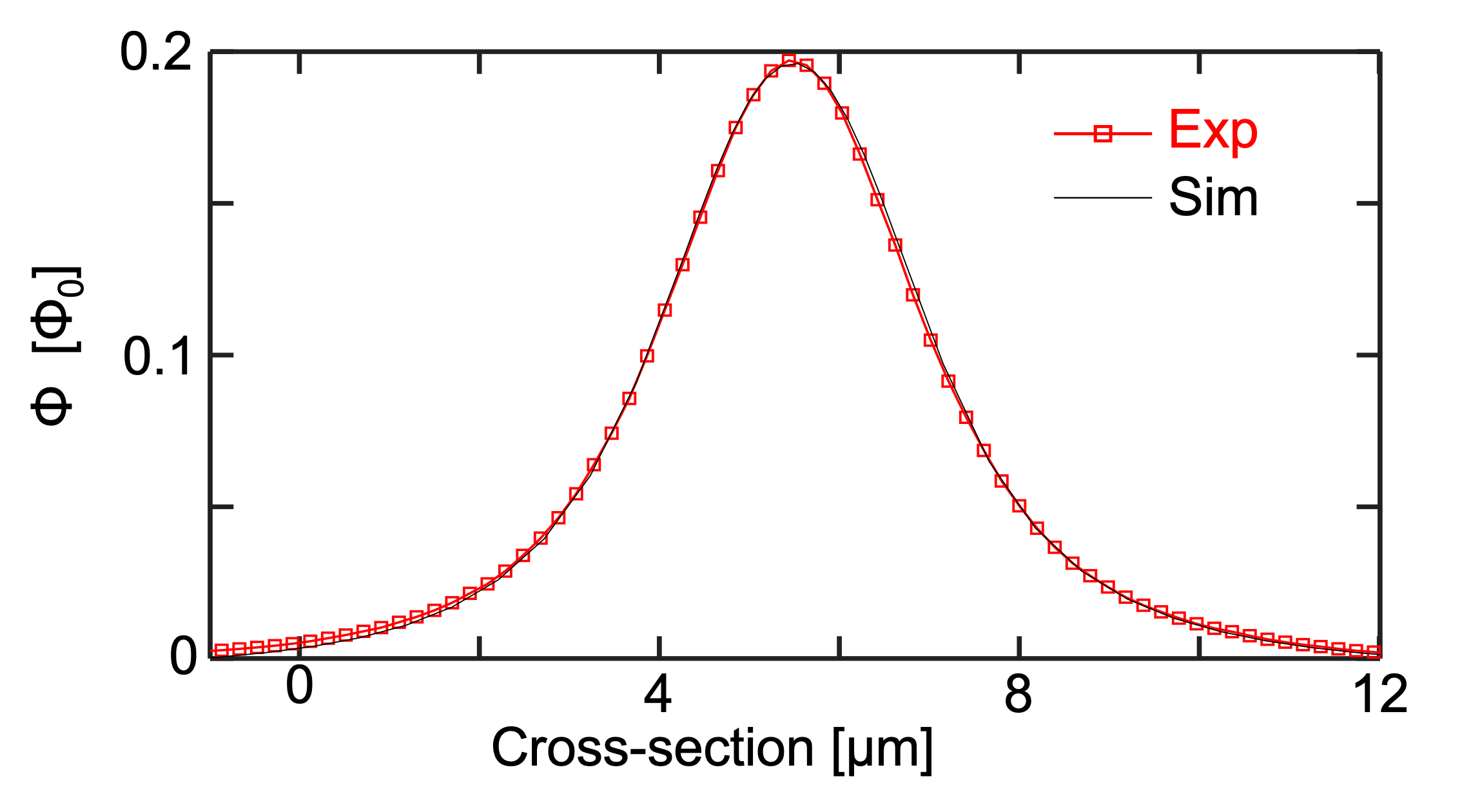}
\caption{\label{fig:1Phi0vortexfit} Cross-section of magnetic field profile of an isolated vortex measured in CaSb$_2$ at 1.5 K [Fig.~1(a)] and monopole model fitting. Scan height is $\sim$1150~nm. The best-fit parameter for the penetration depth is 210 nm, which is consistent with the susceptibility measurement results (170–250~nm at 1.5~K).
}
\end{figure}

\begin{figure}[htb]
\includegraphics*[width=12cm]{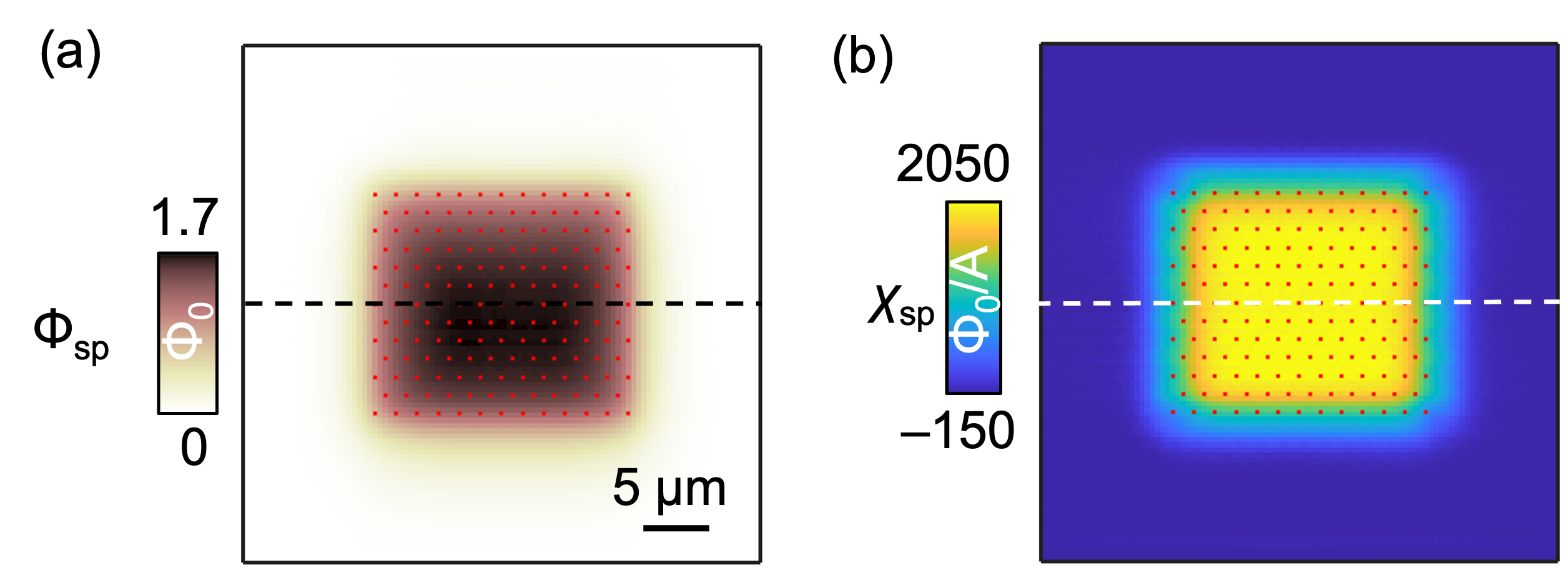}
\caption{\label{fig:largeclusters}Simulation of magnetic fields and dynamics of a triangular-lattice vortex cluster in CaSb$_2$ at 1.5~K.
(a) Simulated magnetic-flux map for lattice constant $a=1.65~\mu$m. The cluster is constructed by linearly superposing the measured flux scan of an isolated 1$\Phi_0$ vortex at 1.5 K [Fig.~1(a)]; $a$ is determined by fitting to the experimental linecut in Fig.~1(f).
(b) Corresponding simulated magnetic-susceptibility map, obtained by linearly superposing the measured isolated-vortex susceptibility scan at 1.5 K [Fig.~1(b)].
    }
\end{figure}

\begin{figure}[htb]
\includegraphics*[width=12cm]{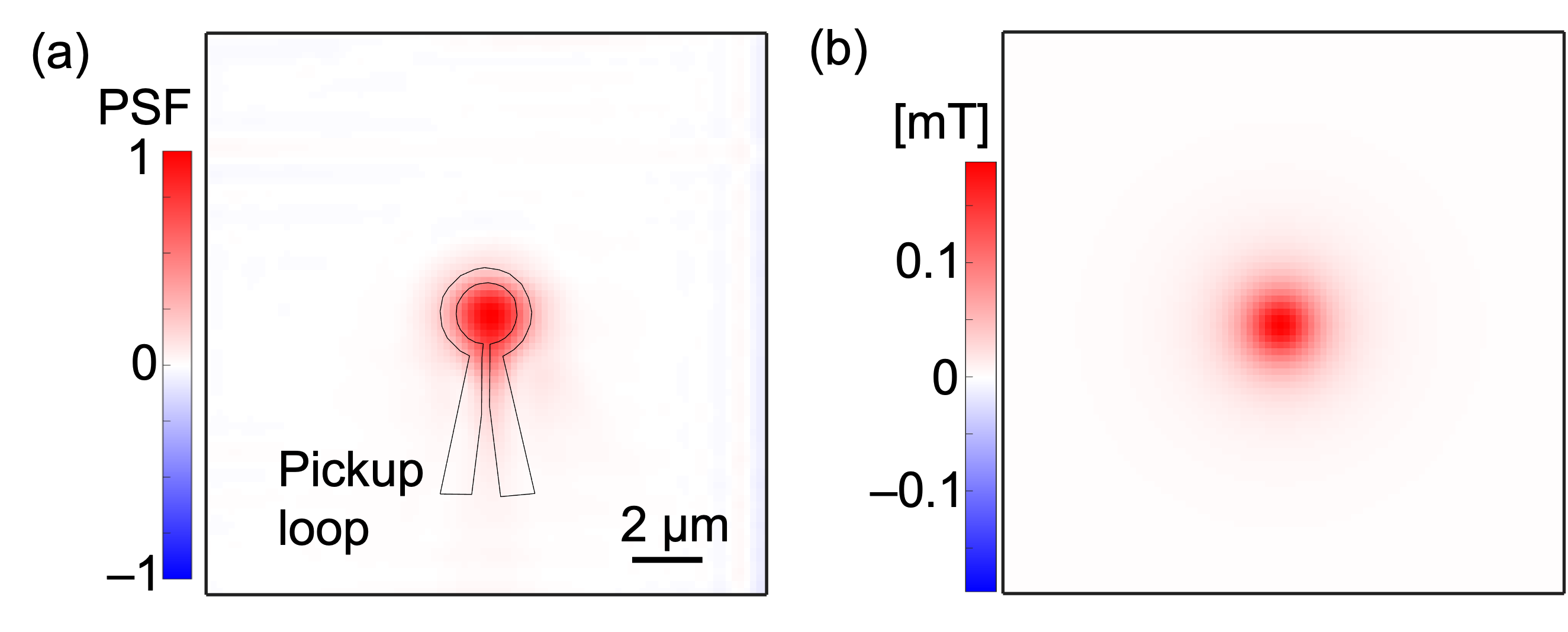}
\caption{\label{fig:PSF} SQUID point spread function calculated from an isolated vortex field profile.
(a) SQUID point spread function (PSF) calculated by using Fig.~1(a) and the $1\Phi_0$ monopole model with the scan height of $\sim$1150~nm and the penetration depth of 170~nm (obtained by Fig.~A1). The tail originates from the shield structure of the pickup loop. The maximum of PSF is normalized so that the integral of PSF corresponds to the effective area of the pickup loop, $\sim$5.2~ \textmu m$^2$. (b) Estimated field profile by using (a) and Fig.~1(a). The integral yields a total magnetic flux of 0.97~$\Phi_0$.
    }
\end{figure}

\begin{figure}[htb]
\includegraphics*[width=17cm]{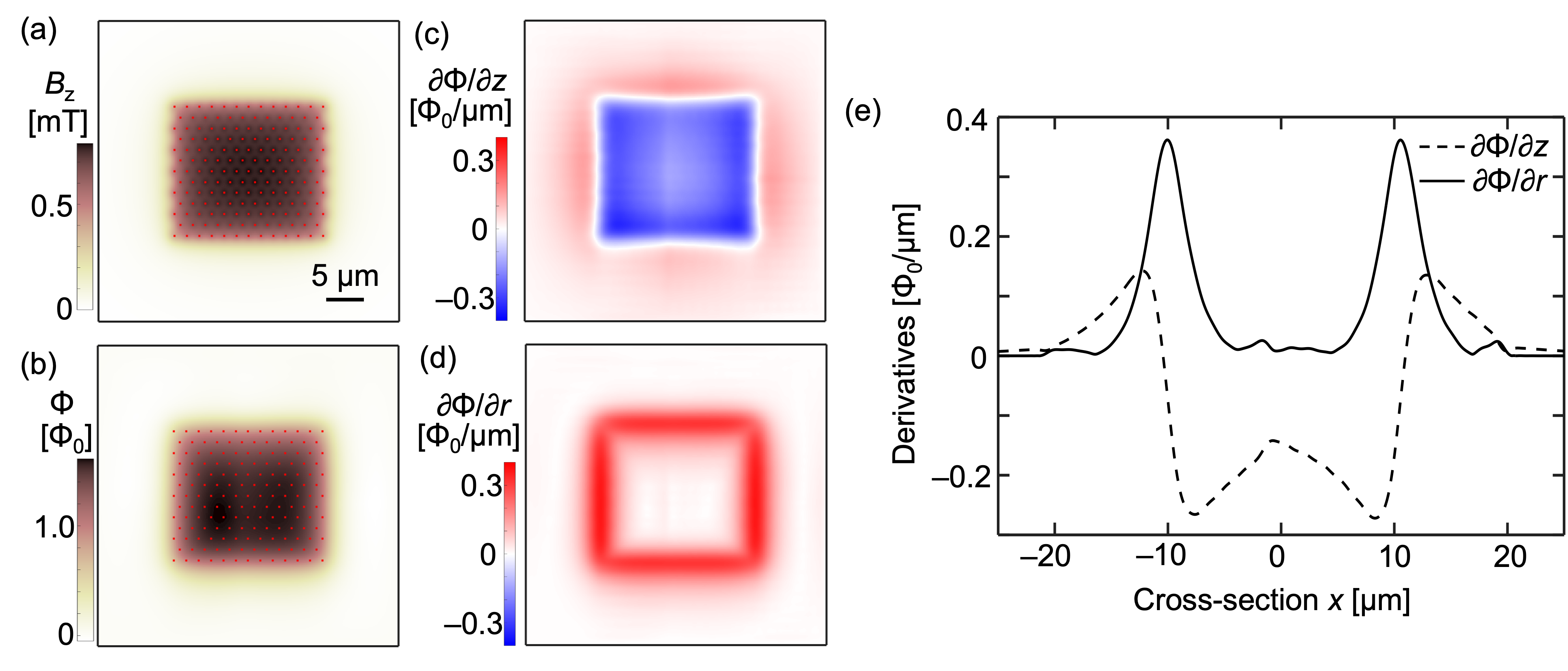}
\caption{\label{fig:PSF_derivatives} Simulated magnetic fields of a triangular vortex cluster in CaSb$_2$ at 1.5 K.
(a) $B_z$ profile computed via the monopole model ($a$=1.65~\textmu m, scan height 1150~nm , penetration depth 170~nm). 
(b) Corresponding magnetic flux $\Phi$ through the SQUID pickup loop using the PSF[Fig.~S3(a)]. 
(c,d) Derivatives of $\Phi$ along the out-of-plane ($z$) and in-plane ($r=\sqrt{x^2+y^2}$) directions. Here $\Delta z = 100$~nm and $\Delta r = 100$~nm. 
(e) Cross-section of derivatives $\partial \Phi/\partial R (R=x,y,z)$ obtained at the cluster center in (c,d).
    }
\end{figure}

\begin{figure}[htb]
\includegraphics*[width=4.5cm]{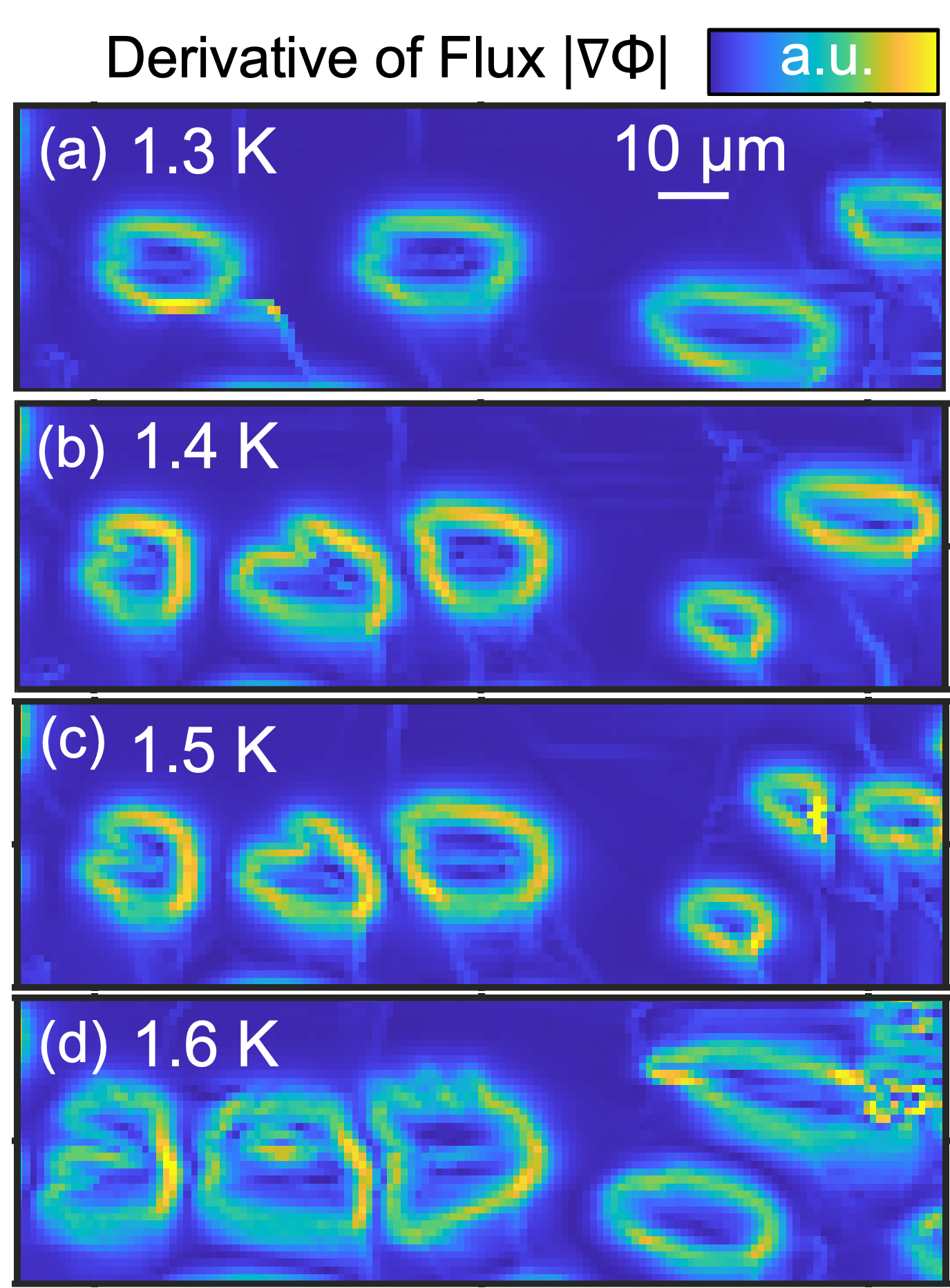}
\caption{\label{fig:T-dep_dphi} Temperature dependence of the derivative of flux $|\nabla\Phi|$  with vortex clusters in CaSb$_2$, calculated from the magnetic flux in Fig. 3 (a-d), only reproduces the peak structure along the cluster outline of the susceptibility in Fig.~3 (e-h).
    }
\end{figure}

\begin{figure}[tb]
\begin{center}
\includegraphics*[width=13cm]{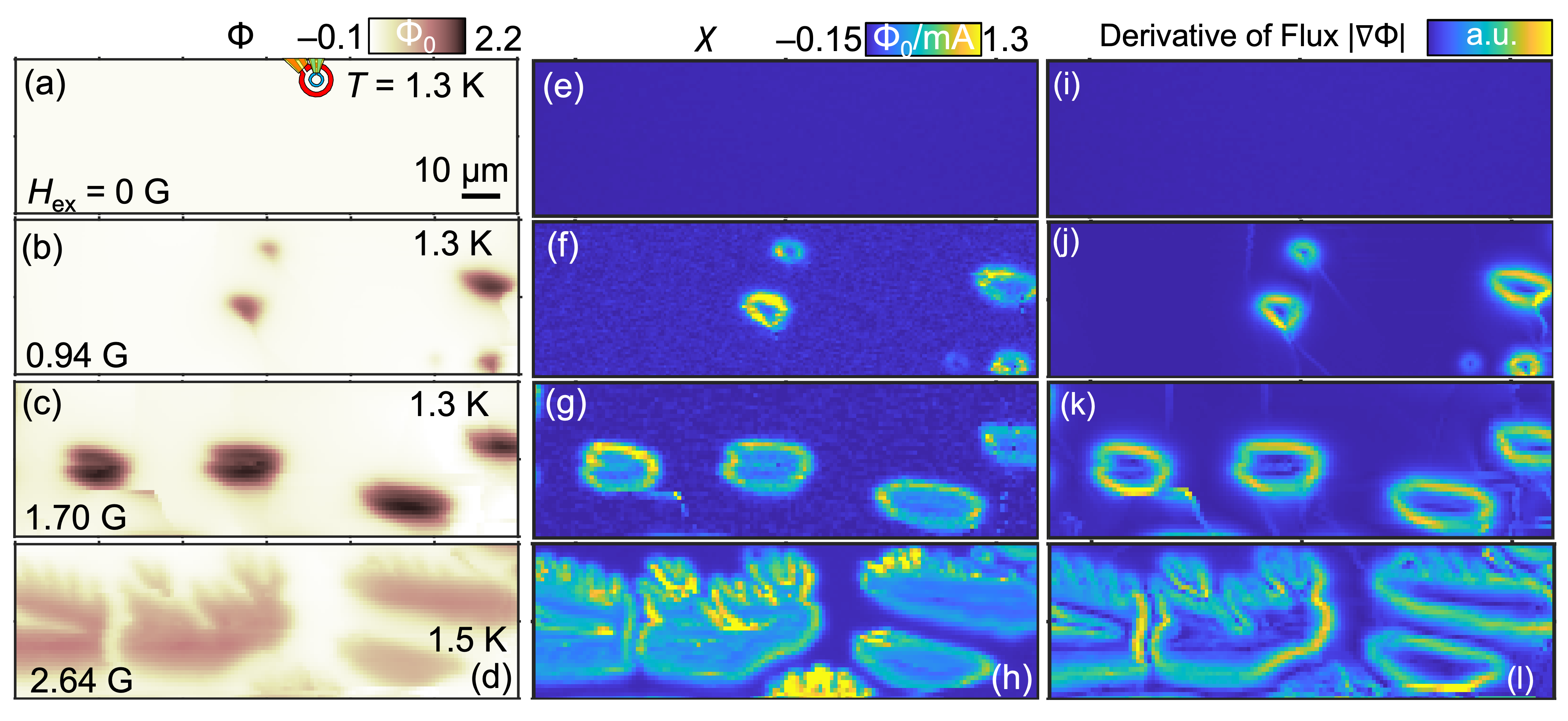}
\caption{\label{fig:Fcool} 
Morphological diversity of vortex clusters in CaSb$_2$ under various field-cooling conditions.
(a–d) Magnetometry, (e–h) susceptometry, and (i-l) the derivative of flux images show that field-cooled vortex clusters can exhibit circular, elongated, or fragmented morphologies. Despite this variety, all clusters consistently exhibit enhanced susceptibility at their boundaries, indicating robust collective dynamics. The Inset shows the schematic of the SQUID sensor at the actual scale.
}
\end{center}
\end{figure}

\end{document}